\documentclass[a4paper,12pt]{article}
\usepackage{amsmath}
\usepackage[pdftex]{graphicx}
\usepackage{color}

\newcommand{\Vect}[1]{\boldsymbol{\mathrm{#1}}{}}
\def\({\left(}
\def\){\right)}
\def\DS{\displaystyle}
\def\av#1{\Bigl\langle{#1}\Bigr\rangle}
\def\be#1{\begin{equation}\label{#1}}
\def\ee{\end{equation}}
\def\eq#1{(\ref{#1})}
\def\edited#1{{ \textcolor{black}{#1}}}

\def\zz{\Vect{z}}
\def\xx{\Vect{x}}
\def\rr{\Vect{r}}
\def\Sa{{\cal S}_{\alpha}}
\def\Rs{\Vect{{\cal R}}}
\def\suma{\sum_{\alpha}}
\def\Ds{{\cal L}}
\def\Dxs{{\cal L}_x}
\def\Dys{{\cal L}_y}
\def\kaph{\hat{\kappa}}
\def\ba{b_{\alpha}}
\def\kk{\Vect{k}}
\def\yy{\Vect{y}}

\def\ve{\Vect{e}}
\def\va{\Vect{a}_{\alpha}}

\def\paptit#1{}

\begin{document}

\baselineskip 16pt
\title{Fast and slow thermal processes in harmonic scalar lattices}
\author{V A Kuzkin
               \and A M Krivtsov\footnote{Peter the Great Saint Petersburg
               Polytechnical University,
              Polytechnicheskaya st. 29, Saint Petersburg, Russia;
              Institute for Problems in Mechanical Engineering RAS, Bolshoy pr.
              V.O. 61, Saint Petersburg, Russia;
              e-mails: kuzkinva@gmail.com (V.A. Kuzkin), akrivtsov@bk.ru (A.M.
              Krivtsov)}
}
\maketitle



\def\sect#1{\section{#1}}


\begin{abstract}
An approach for analytical description of thermal processes in harmonic
lattices is presented. We cover longitudinal and transverse
vibrations of chains and out-of-plane vibrations of two-dimensional lattices
with interactions of an arbitrary number of neighbors. Motion of each particle is governed by a single scalar equation and therefore the notion ``scalar lattice'' is used. Evolution of initial temperature field in \edited{an infinite} lattice is investigated. An exact equation describing the evolution is derived. Continualization of this equation with respect to spatial coordinates is carried out. The resulting continuum equation is solved analytically. The solution shows that the kinetic temperature is represented \edited{as the sum of two terms, one describing short time behavior, the other large time behavior}. At short times, the temperature performs high-frequency oscillations  caused by
redistribution of energy among kinetic and potential forms~(fast process).
Characteristic time of this process is of order of ten periods of atomic vibrations. At large times, changes of the temperature are caused by ballistic heat transfer~(slow process). The temperature field is represented
as a superposition of waves having the shape of initial temperature
distribution and propagating with group velocities dependent on the wave
vector. Expressions describing fast and slow processes are invariant with respect
to substitution~$t$ by $-t$. However examples considered in the paper demonstrate that these processes are irreversible. Numerical simulations show that presented theory describes the evolution of temperature field at short and large time scales with high accuracy.
\end{abstract}

\edited{
}\\
\edited{
}


\section{Introduction}
At macrolevel, heat propagation is usually diffusive and well-described by the
Fourier law. The law assumes linear dependence between the heat flux and
temperature gradient with proportionality coefficient refereed to as the heat
conductivity. Phonon theory relates the heat conductivity coefficient with the
phonon mean free path~\cite{Pierls, Ziman phonon}.
It is assumed that the Fourier law is valid if the mean free path is much
smaller than characteristic size of the system. At micro- and nanolevel, this
condition may be violated. In particular, it is shown experimentally that the
mean free path can be as large as several microns~\cite{mfp SiGe}. In this
case, the heat transport is ballistic~\cite{mfp SiGe, Goodson review exp, Liu review exp,
Chang review exp, Chen ballistic, Pumarol ballistic, Cenian ballistic} and can not be described by the Fourier law. In particular,
the effective heat conductivity is size-dependent~\cite{Goodson review exp, Liu
review exp, Chang review exp} and it can not be regarded as a material
constant. This phenomenon leads to a variety of practical applications of heat
transport in micro- and nanosystems~(see e.g. review paper~\cite{Shi review
applications}). On the other hand, derivation of macroscopic heat transport
equations from lattice dynamics equations is a serious challenge for
theoreticians~\cite{Lebowitz 2000}.

One of the convenient models for investigation of heat transport in solids is a
harmonic crystal.  Heat transport in harmonic crystals is usually investigated
in a {\it steady-state regime}. Stationary temperature distribution  between two
reservoirs with different temperatures\footnote{Here and below kinetic
temperature is considered.} is considered. For example, in a pioneering work by
Reider, Lebowitz and Lieb~\cite{Lebowitz1967}, anomalies of the heat transport
in one-dimensional harmonic chain with nearest neighbor interactions are
demonstrated and an analytical solution of the steady heat transport
problem is derived. The solution shows that thermal resistance of the
chain\footnote{Thermal resistance is the inverse of the heat conductivity.} is
independent on its length. Therefore the effective heat conductivity diverges
with length and the Fourier law is not applicable.

Anomalous heat transport is also observed in more complicated harmonic systems.
Generalization of results obtained in paper~\cite{Lebowitz1967}  for the
multidimensional case is carried out in papers~\cite{Allen1969, Nakazawa1970}.
Harmonic chains with alternating masses are considered in paper~\cite{Dhar
altern mass}. The effect of disorder on heat transport in harmonic crystals is
studied in papers~\cite{Lee_Dhar_2005, Dhar2010disorder, Gonzalez2010, Gonzalez2015}. \edited{The influence of conservative
bulk noises on nonequilibrium steady state is investigated in
papers~\cite{Politi2009, Politi2010, Dhar2011 bulk noise, Bernardin2012}.}  Solution of the steady-state
problem for an arbitrary harmonic network is obtained in paper~\cite{Freitas
Paz 2014}.
Specific feature of the stationary heat transfer problem considered in the
above mentioned papers is that the results strongly depend on the type of
thermostat~\cite{DharSaito2016, Lepri 2003, Bonetto2004}. For example, in
paper~\cite{Bonetto2004} it is shown that a specific choice of the thermostat
leads to Fourier heat conduction in harmonic crystals. Note that the influence
of thermostat is also observed in nonlinear systems~\cite{Hoover diff
thermostats}.

\edited{The present paper focuses on {\it unsteady} thermal processes. Evolution of initial temperature field in an infinite lattice is considered.} It allows us to
investigate properties of the lattice, rather than properties of the thermostat. We consider the initial conditions typical for molecular dynamics simulations of the heat transfer~\cite{Allenbook, Tsai nonstat, Hoover heat cond, JonesZimmerman, Jones, Gendelman 2010 nonstat, Krivtsov_LeZakharov, KosevichSavin, PiazzaLepri}.
Initially, particles have random velocities corresponding to the initial temperature field. Initial displacements are equal to zero. In this case, initial kinetic and potential energies are not equal. \edited{Motion of particles leads to redistribution of energy between kinetic and potential forms\footnote{\edited{Since the total energy is conserved, then the kinetic energy is converted to potential energy.}}. After some time the energies become equal as predicted by the virial theorem~\cite{Hoover_stat_phys}. However the theorem does not describe the transient process.} This process is important in particular because it determines the short time behavior of the kinetic temperature.
Analytical description of the transient process is reported only for several particular systems~\cite{Krivtsov 2014 DAN, Babenkov 2016, Kuzkin_FTT}. At large time scale, kinetic and potential energies are practically equal and changes of kinetic temperature are caused by the energy transport. Rigorous mathematical description of large time behavior of energy density for harmonic lattices in continuum limit is presented in papers~\cite{Melike, Lukkarinen Spohn 2007, Dudnikova Spohn transport eq,  Harris 2008 SIAM, Lukkarinen 2016}.  Short time behavior of kinetic temperature is not considered in these works.

The main goal of the present paper is to develop an approach for analytical description of short and large time behavior of the kinetic temperature. A wide class of one- and two-dimensional lattices with interactions of an arbitrary number of neighbors and harmonic on-site potential is considered. The approach is based on analysis of velocity covariances for all pairs of particles. A deterministic equation exactly describing the evolution of temperature field is derived~(section~\ref{sect fast slow}). Continualization of this equation with respect to  spatial coordinates is carried out~(section~\ref{sect-contin}). The resulting continuum equation is solved analytically~(section~\ref{sect-fast-slow-gen}). The expression for the temperature field, valid at both short and large time scales, is obtained.
At short times, it describes oscillations of temperature caused by equilibration of kinetic and potential energies~(fast process). From a practical viewpoint, the description of the fast process is important for modeling of fempto- and attosecond laser excitation~\cite{Corkum attosecond, Petrov Fortov, Poletkin, Balandin graphene}. At large time scale, the expression describes the ballistic heat transfer~(slow process)\footnote{The term ``ballistic heat transfer'' is also used in papers~\cite{mfp SiGe, Goodson review exp, Liu review exp,
Chang review exp, Chen ballistic, Pumarol ballistic}.}. Large time behavior of  the solution is consistent with results obtained in papers~\cite{Melike, Lukkarinen Spohn 2007, Dudnikova Spohn transport eq,  Harris 2008 SIAM, Lukkarinen 2016}. Additionally, the present work contains detailed analysis of several physically
important cases, such as, oscillations of temperature in uniformly heated lattice~(section~\ref{fastsq}),  contact of hot and cold
half-spaces~(sections~\ref{sect divergence heat conductivity}, \ref{sect
Heaviside}) and irreversible decay of sinusoidal temperature
distribution~(section~\ref{sect_sinus}). Analytical results are supported by
numerical simulations.


\section{Equations of motion and initial conditions}\label{sect EM ICS}
Consider an infinite harmonic lattice with simple structure\footnote{Unit cell
of the lattice contains only one particle.} in $d$-dimensional space,
where~$d=1,2$. Each particle has one degree of freedom, i.e. particles move
along parallel lines. Displacement of a particle is described by the scalar
function~$u(\xx)$, where~$\xx$ is the radius vector of the particle in the
undeformed state.  Therefore the notion {\it ``scalar
lattice''}~\cite{Melike, Harris 2008 SIAM, Gendelman 2016, Nishiguchi et al, Mishuris}  is used.

Each particle interacts with neighbors numbered by index~$\alpha$.
 Vectors~$\va$, connecting the particle with its  neighbors, satisfy
 relation\footnote{Then~$\Vect{a}_0 = 0$.}
\be{}
 \va = -\Vect{a}_{-\alpha}.
\ee
Since the lattice is harmonic, then the total force acting on the particle
is a linear combination of displacements of the neighboring particles.
Therefore the equations of motion have the form
\be{EM}
  \begin{array}{l}
   \DS  \ddot{u}\(\xx\) = \Ds u\(\xx\),
    \qquad
   \Ds u\(\xx\) = \omega_*^2 \sum_{\alpha} \ba u(\xx + \va), \quad \ba =
   b_{-\alpha},
  \end{array}
\ee
where~$\omega_*$ is a characteristic frequency~(see e.g. formulas~\eq{om 1D},
\eq{ba square});
$\Ds$ is a linear difference operator\footnote{From mathematical point of view,
formula~\eq{EM} is a differential-difference equation or an infinite set of
coupled ODE's of the second order.}.

Equations of motion~\eq{EM} cover a variety of one- and two-dimensional
lattices. For example, the simplest one-dimensional lattice described by
equation~\eq{EM} is a chain with nearest-neighbor interactions. In this case
\be{om 1D}
\begin{array}{l}
\Ds u\(\xx\) = \omega_*^2 \Bigl( u(\xx + \Vect{a}_{1}) - 2 u(\xx) + u(\xx +
\Vect{a}_{-1})\Bigr) \quad \Rightarrow
\\[4mm]
\Rightarrow  \quad \omega_* = \sqrt{\frac{C}{M}},
\qquad
\quad \Vect{a}_{\pm 1} = \pm  a \Vect{i},
\qquad b_{\pm 1}=1,
\quad b_0 = -2,
\end{array}
\ee
where~$\Vect{i}$ is a unit vector directed along the chain; $a$ is an
equilibrium distance between neighboring particles; $C$ is bond stiffness; $M$
is particle mass. Note that equation~\eq{om 1D} with appropriate choice
of~$\omega_*$ also describes linearized transverse vibrations of a stretched
chain with pair interactions~\cite{Liu harm chain trans, Kuzkin2015chain}.
Transverse vibrations of a chain with angular
interactions~\cite{Wang 2004 mult chain transverse, Kuzkin2010} are also
described by equation~\eq{EM}.
In this case
\be{om 1D transverse}
\begin{array}{l}
\Ds u\(\xx\) = -\omega_*^2 \Bigl( u(\xx + \Vect{a}_{2}) -4u(\xx + \Vect{a}_{1})
+ 6 u(\xx) -4u(\xx + \Vect{a}_{-1})+ u(\xx + \Vect{a}_{-2})\Bigr) \quad
\Rightarrow
\\[4mm]
\Rightarrow  \quad \omega_* = \sqrt{\frac{C_a}{Ma^2}},
\qquad
\Vect{a}_{\pm 1} = \pm  a \Vect{i},
\qquad
\Vect{a}_{\pm 2} = \pm 2a \Vect{i},
\qquad
b_{0} = -6,
\quad
 b_{\pm 1} = 4,
\quad
b_{\pm 2} = -1,
\end{array}
\ee
where~$C_a$ is  stiffness of the angular spring. This model can be
used, for example, for description of ballistic heat transfer in carbine. \edited{It can also be considered as a coarse-grained model for nanowires~\cite{mfp SiGe} or diamond nanothreads~\cite{Xu nanothread}.}

The simplest two-dimensional system described by equations of motion~\eq{EM} is
a stretched square lattice with nearest-neighbor interactions performing
out-of-plane vibrations.
In this case
\be{ba square}
\begin{array}{l}
\Ds u\(\xx\) = \omega_*^2 \Bigl( u(\xx + \Vect{a}_{1}) + u(\xx + \Vect{a}_{2})
- 4 u(\xx) + u(\xx + \Vect{a}_{-1}) + u(\xx + \Vect{a}_{-2})\Bigr) \quad
\Rightarrow
\\[4mm]
\Rightarrow
\quad
\omega_* = \sqrt{\frac{F}{M a}},
\quad \Vect{a}_{\pm1} =  \pm a \Vect{i},
\quad \Vect{a}_{\pm 2} =  \pm a \Vect{j},
\quad b_{\pm 1}=b_{\pm 2}=1, \quad b_0 = -4,
\end{array}
\ee
where $\Vect{i}$, $\Vect{j}$ are orthogonal unit vectors;
$F$ is the magnitude of stretching force  in equilibrium. This lattice is
considered in detail in section~\ref{sect square}. \edited{Two-dimensional scalar lattices can be considered as simplest models for out-of-plane~(transverse) vibrations of 2D materials such as graphene~\cite{Balandin graphene rev, Xie graphene, Berinskii gr, Xu boron nitride}, molybdenum disulphide~\cite{Berinskii mol, DmitrievMO}, boron nitride~\cite{Xu boron nitride}, etc.}

{\bf Remark.} In general, an appropriate choice of parameters~$\va$ and $\ba$ in~\eq{EM}
allows to consider linearized vibrations of one- and two-dimensional scalar
lattices. Pair and multibody  interactions with an arbitrary number of neighbors
and harmonic on-site potential can be considered.\footnote{Equation~\eq{EM} also covers some systems with torque interactions~\cite{Kuzkin2012, Mishuris2013a}. For example, a chain consisting of connected rigid bodies with fixed translational degrees of freedom~\cite{Dievich} is described by equation~\eq{EM}. Out-of-plane
vibrations of 2D lattices can also be
considered provided that rotational degrees of freedom are fixed.}

We consider the following {\it stochastic} initial conditions typical for molecular dynamics
simulations:
\be{IC part}
   u(\xx) = 0, \qquad  v(\xx) = v_0\(\xx\),
\ee
where~$v = \dot{u}$; initial velocities~$v_0\(\xx\)$ are uncorrelated, centered
 random numbers with zero mean. Initial conditions~\eq{IC part} correspond to some instantaneous distribution
of kinetic temperature in a lattice.

Note that no assumptions about  distribution
function for velocities are made. Evolution of the distribution function and
its convergence to the Gaussian distribution are discussed e.g. in
papers~\cite{Dudnikova2003, Dudnikova2004, Kozlov Treschev}.

Equations of motion~\eq{EM} with initial conditions~\eq{IC part} can be solved
analytically. The solution yields random particle displacements and
velocities.
In contrast, description of macroscopic thermal processes
usually focuses on statistical characteristics
such as a kinetic temperature. An equation exactly describing the evolution
temperature field is derived in the following section.

\section{Covariances of velocities. Kinetic temperature} \label{sect fast
slow}
\label{correlations}
In the present section, we derive an equation for covariances of particle
velocities. Solution of this equation {\it exactly} describes evolution of kinetic
temperature.

A covariance of velocities for
particles with radius-vectors~$\xx$ and $\yy$ is defined as
\be{xi ka}
 \kappa(\xx,\yy) = \av{v(\xx) v(\yy)}.
\ee
Here and below angle brackets~$\av{...}$ stands for mathematical
expectation\footnote{In numerical simulations, the mathematical expectation can
be approximated by an average over realizations with different initial
conditions~(see e.g. section~\ref{sect Heaviside}).}. The number of
covariances~\eq{xi ka} is equal to the number of different particle pairs in
the lattice. The velocity covariance is related to kinetic temperature~$T$ by
the following formula:
\be{kin temp}
 k_B T(\xx) = M \av{v(\xx)^2} = M \kappa|_{\xx=\yy},
\ee
where~$k_B$ is the Boltzmann constant.

Differentiation of covariances~\eq{xi ka} with respect to time taking into
account equations of motion~\eq{EM}, yields the following equation~(see
 appendix~\ref{eqcovariances} for more details):
\be{4order}
\begin{array}{l}
\DS\ddddot{\kappa} - 2\(\Dxs + \Dys\)\ddot{\kappa}  + \(\Dxs-\Dys\)^2\kappa =
0,
\end{array}
\ee
where
\be{}
\DS \Dxs \kappa  = \omega_*^2 \sum_{\alpha} \ba \kappa(\xx + \va,\yy),
\quad
\Dys \kappa = \omega_*^2 \sum_{\alpha} \ba \kappa(\xx, \yy+ \va).
\ee
Equation~\eq{4order} {\it exactly}  describes the evolution of
temperature field in {\it any} harmonic scalar lattice.

Initial conditions for equation~\eq{4order}, corresponding to initial
conditions
for particles~\eq{IC part}, have the form:
\be{IC kappa}
\begin{array}{l}
\DS \kappa = \frac{k_B }{M}T_0(\xx)\delta_D(\xx-\yy),
        \quad  \dot{\kappa} = 0,
        \quad  \ddot{\kappa} = \frac{k_B}{M}\(\Dxs + \Dys\)\( T_0(\xx)\delta_D(\xx-\yy)\),
        \quad  \dddot{\kappa} = 0,
        \\[4mm]
\DS k_B T_0(\xx) = M \av{v_0(\xx)^2},
\end{array}
\ee
where $T_0(\xx)$ is the spatial distribution of initial temperature; function~$\delta_D(\xx-\yy)$ is equal
to unity  for~$\xx=\yy$ and it is equal to zero for~$\xx \neq \yy$.

Thus an exact deterministic equation~\eq{4order} is obtained for the
stochastic thermal problem. In the following sections, we construct  solutions of equation~\eq{4order} in continuum limit.

\edited{{\bf Remark.} Analysis of covariances can also be used for description of thermal processes in harmonic chains with a conservative noise~\cite{Politi2009, Politi2010}. In this case, it is not sufficient to consider covariances of velocities. Equations for covariances of displacements and cross-covariances of velocities and displacements should be added in order to obtain closed system of equations.}

\section{Continualization}\label{sect-contin}
In the present section, we simplify equation~\eq{4order} using continualization
with respect to spatial variable~\cite{Krivtsov DAN 2015, Born}.
We introduce new variables:
\be{kappa z}
(\xx,\yy) \rightarrow \(\rr,\,\xx-\yy\), \qquad \rr = \frac{\xx+\yy}{2}.
\ee
From now on, covariance of velocities is represented in the form~$\kappa(\rr,
\xx-\yy)$.
Continualization of equation~\eq{4order} is carried out with respect to spatial
variable~$\rr$.

We assume that function~$\kappa$ is slowly changing with~$\rr$ at distances of
order of~$|\va|$. Then operators~$\Dxs, \Dys$ can be approximated by the power
series expansion with respect
to~$\va$~(see. appendix~\ref{long wave}):
\be{series}
\begin{array}{l}
\Dxs \approx \Ds +  \Rs \cdot \nabla, \qquad \Dys  \approx \Ds - \Rs \cdot
\nabla,
\\[4mm]
\DS
\Ds = \omega_*^2 \suma \ba \Sa,
\quad
\Rs  = \frac{\omega_*^2}{2}  \suma \va\ba\Sa,
\quad
\Sa \kappa = \kappa\(\rr, \xx-\yy + \va\),
\end{array}
\ee
where~$\nabla = \frac{\partial}{\partial \rr}$ is nabla-operator.
Substitution of formula~\eq{series} into~\eq{4order}, \eq{IC kappa} yields equation
\be{4order cont}
\ddddot{\kappa} - 4\Ds\ddot{\kappa}  + 4\(\Rs \cdot \nabla\)^2\kappa = 0
\ee
with initial conditions
\be{IC kappa cont}
 \kappa = \frac{k_B }{M}T_0(\rr)\delta_D(\xx-\yy),
        \quad  \dot{\kappa} = 0,
        \quad  \ddot{\kappa} = 2\frac{k_B }{M}T_0(\rr)\Ds\delta_D(\xx-\yy),
        \quad  \dddot{\kappa} = 0.
\ee

Equation~\eq{4order cont} describes, in particular, the evolution of
 temperature field in continuum limit. The equation
is differential with respect to continuum
variables~$\rr$, $t$ and difference with respect to discrete-valued
variable~$\xx-\yy$.

{\bf Remark.} In the case of uniform distribution of initial temperature~($T_0 = {\rm const}$) covariances of velocities exactly satisfy the following equation:
\be{4order exact uniform}
 \ddddot{\kappa} - 4\Ds\ddot{\kappa} = 0.
\ee

Solutions of equations~\eq{4order cont} and \eq{4order exact uniform} are derived below.

\section{Analytical solution: fast and slow thermal processes}\label{sect-fast-slow-gen}
In the present section, we solve equation~\eq{4order cont} and obtain the expression describing the evolution of the temperature field.

The solution is constructed using the discrete Fourier transform~(see
Appendix~\ref{c_f general} for definition). Since lattices with simple structure are considered, then vectors~$\xx-\yy$ are represented in the form
\be{}
 \DS    \xx-\yy = a \sum_{j=1}^d z_j \ve_j,
\ee
where~$\ve_j$, $j=1,..,d$ are unit vectors directed along basis vectors of the
lattice;  $d$  is space dimensionality; $a$ is an equilibrium distance;  $z_j$
are integer numbers.

Applying the discrete Fourier transform to equation~\eq{4order cont} with respect
to~$z_j$, yields
\be{4order cont F}
\begin{array}{l}
\ddddot{\kaph} + 4\omega^2\ddot{\kaph}  - 4\omega^2\(\Vect{c} \cdot \nabla\)^2\kaph = 0.
\end{array}
\ee
Here~$\omega(\kk)$ is the dispersion relation for the lattice,
$\kk$ is the wave vector:
\be{disp rel}
   \omega^2(\kk) = -\omega_*^2 \(b_0 + 2\sum_{\alpha > 0} b_{\alpha} \cos\(\kk
     \cdot \va\)\),
     \qquad
     \kk = \frac{1}{a}\sum_{j=1}^d p_j\tilde{\ve}_j,
\ee
where $\tilde{\ve}_j$
 are vectors of
the reciprocal basis.\footnote{Vectors of the
reciprocal basis~$\tilde{\ve}_j$ are defined
as~$\ve_j\cdot\tilde{\ve}_k = 1$
for $j=k$ and $\ve_j\cdot\tilde{\ve}_k =0$ for~$j\neq k$.}
Vector~$\Vect{c}$ coincides with vector of group velocity
for the lattice:
\be{group vel general}
 \Vect{c}
 =
 \frac{{\rm d} \omega}{{\rm d} \kk}
 =
 \frac{\DS \omega_*  \sum_{\alpha>0} \ba \va \sin\(\kk \cdot \va\)}{\DS
 \sqrt{-b_0 - 2\sum_{\alpha>0} \ba\cos\(\kk \cdot \va\)}},
\ee
Formulas~\eq{disp rel}, \eq{group vel general} are derived in  appendix~\ref{c_f general}.

To the accuracy of term~$\left(\Vect{c} \cdot \nabla\right)^2\ddot{\kaph}$, neglected during the continualization,
equation~\eq{4order cont F} is factorized
\be{factorized}
 \(\frac{\partial^2}{\partial t^2} +  4\omega^2\)\( \frac{\partial^2}{\partial t^2} - \(\Vect{c} \cdot \nabla\)^2\)\kaph = 0.
\ee
Solution of equation~\eq{factorized} is equal to sum of
solutions of the following equations
\be{Fast}
\ddot{\kaph} +  4\omega^2 \kaph = 0,
\ee
\be{Slow}
 \ddot{\kaph} - \(\Vect{c} \cdot \nabla\)^2 \kaph = 0.
\ee
Solving equations~\eq{Fast}, \eq{Slow} with initial conditions~\eq{IC kappa cont}
and applying the inverse discrete Fourier transform, yields
\be{sol fast slow}
\DS T = T_F + T_S,
\ee
\be{fast gen}
 \DS T_F = \frac{T_0(\rr)}{2(2\pi)^d} \int_{-\pi}^{\pi} \cos\bigl(2\omega t\bigr){\rm d} p_1... {\rm d} p_d,
\ee
\be{sol slow}
 \DS
 T_S =
 \frac{1}{4(2\pi)^d}\int_{-\pi}^{\pi}
 \Bigl(T_0(\rr+\Vect{c} t) + T_0(\rr-\Vect{c} t)\Bigr){\rm d} p_1... {\rm d} p_d.
\ee
Formulas~\eq{sol fast slow}, \eq{fast gen}, \eq{sol slow} describe the behavior  of the kinetic temperature. \edited{It is seen that the temperature is represented as a sum of two terms. The first term, $T_F$, describes short time behavior of the kinetic temperature, while the second term, $T_S$, describes large time behavior.}

At short times, the kinetic temperature performs high-frequency oscillations caused by redistribution of energy among kinetic and potential
forms~(fast process)\footnote{\edited{Note that in lattices with several degrees of freedom per unit cell there is an additional fast process. It is caused by redistribution of energy among the degrees of freedom. See e.g. paper~\cite{Kuzkin_FTT}.}}. According to formula~\eq{fast gen}, these oscillations in different spatial points are independent.   Integrand in formula~\eq{fast gen}  changes sign
and oscillates with frequency proportional to time. Therefore~$T_F$
tends to zero\footnote{Rigorous proof of this fact is beyond the scope of the
present paper. Investigation of integrals of this type can be carried out using
asymptotic methods~\cite{Wong}.}, while temperature tends to~$T_S$. To our knowledge, general analytical description
of this fast transient process for scalar lattices is not
presented in literature.\footnote{Several particular systems, namely one-dimensional chain with nearest-neighbor interactions and two-dimensional triangular lattice, are considered in
papers~\cite{Krivtsov 2014 DAN, Babenkov 2016} and~\cite{Kuzkin_FTT} respectively.} Detailed analysis of the process for the stretched square lattice performing
out-of-plane vibrations is presented in \edited{sections~\ref{fastsq} and \ref{sect_sinus}}.

\edited{{\bf Remark.} In a uniformly heated crystal~($T_0={\rm const}$) formula~\eq{fast gen} is an exact solution. In this case, temperature tends to the stationary value~$T_0/2$. This fact also follows from the virial
theorem~\cite{Hoover_stat_phys}. However in contrast to formula~\eq{fast gen}, the virial theorem does not describe the transition to the stationary state.}

At large time scale, the fast process decays~($T_F \approx 0$) and
changes of the  temperature field are caused by ballistic heat transfer. Then the first term in formula~\eq{sol fast slow} vanishes, i.e. $T \approx T_S$,
where~$T_S$ is defined by formula~\eq{sol slow}.  Formula~\eq{sol slow} shows that at large times, the temperature field is represented as the superposition of waves
traveling with group velocities~$\Vect{c}(\kk)$ and having a shape of initial
temperature distribution~$T_0$.

{\bf Remark.} The latter fact is consistent with results obtained in papers~\cite{Melike, Lukkarinen Spohn 2007, Dudnikova Spohn transport eq,  Harris 2008 SIAM, Lukkarinen 2016} using different formalism. In these works, it is shown that large time behavior of the Wigner function~(``wavenumber resolved'' energy density~\cite{Harris 2008 SIAM}) is governed by the energy transport equation, similar to equation~\eq{Slow}. \edited{In paper~\cite{Spohn 2005}, it is shown that the energy transport equation is a limiting case of the Boltzmann Transport Equation corresponding to zero phonon scattering.
In the present paper, equation~\eq{Slow} is derived and solved for velocity covariances rather than energies. However at large times, the behavior of energy and kinetic temperature is similar, and therefore the results are consistent. Note that short time behavior of temperature is not considered in papers~\cite{Melike, Lukkarinen Spohn 2007, Dudnikova Spohn transport eq,  Harris 2008 SIAM, Lukkarinen 2016, Spohn 2005}.}

{\bf Remark.} Formulas~\eq{fast gen}, \eq{sol slow}
describing thermal processes in scalar lattices are symmetric with respect to time~(invariant with respect to the substitution~$t \rightarrow -t$). However thermal processes are irreversible~(see sections~\ref{fastsq}, \ref{sect_sinus}). \edited{This finding is consistent with results obtained in paper~\cite{Spohn Lebowitz 1977}, where it is shown that locally disturbed infinite harmonic systems return to equilibrium state. The irreversibility is caused by an infinite system size.}

\section{Fundamental solution of ballistic heat transfer problem}

\subsection{One-dimensional chains. Speed of the heat front}
In the present section, we derive the fundamental solution of the heat transfer
problem for one-dimensional chains described by equations~\eq{EM}. Large time behavior of the kinetic temperature is considered~($T\approx T_S$).

Initial distribution of temperature reads
\be{IC delta 1D}
   T_0(x) = A \delta(x),
\ee
where~$\delta$ is the Dirac delta function.
The multiplier~$A$ is introduced in order to obtain solution in proper units.
In this case the solution~\eq{sol slow}
has the form:
\be{kappa1D}
 T_S = \frac{A}{4\pi} \int_{0}^{\pi} \Bigl(\delta\(x-c t\) + \delta\(x+c
 t\)\Bigr) {\rm d} p,
 \qquad
  c
 =
 \frac{\DS \omega_* a \sum_{\alpha>0} \ba \alpha \sin\(\alpha p\)}{\DS
 \sqrt{-b_0 - 2\sum_{\alpha>0} \ba\cos\(\alpha p\)}}.
\ee
The integral is calculated using the identity~\cite{Gelfand}:
\be{ident delta}
 \int \delta\Bigl(\phi(x)\Bigr) \psi(x) {\rm d}x = \sum_j
 \frac{\psi(x_j)}{|\phi'(x_j)|}, \qquad \phi(x_j) = 0.
\ee
Here summation is carried out over real roots, $x_j$, of the equation~$\phi(x) = 0$.
Calculation of the integral~\eq{kappa1D} using identity~\eq{ident delta},
yields the fundamental solution:
\be{delt_1D}
    T_S = \frac{A}{4\pi t} \sum_j \frac{1}{|c'(p_j)|},
    \qquad
    |c(p_j)| = \frac{|x|}{t},
    \qquad
    c'=\frac{{\rm d}c}{{\rm d}p}.
\ee
Here summation is carried out over all real roots, $p_j
\in [-\pi; \pi]$, of the second equation. Function~$c$ is defined by formula~\eq{kappa1D}.
Formula~\eq{delt_1D}  shows that the temperature tends to infinity at extremes
of function~$c(p)$.

Thus formula~\eq{delt_1D} gives the fundamental solution of the heat transfer
problem for one-dimensional chains with interactions of an arbitrary number of neighbors. The general solution corresponding to
the initial temperature distribution~$T_0(x)$ has the form:
\be{sol using delta 1D}
 T_S = \frac{c_*}{4\pi} \sum_j \int_{-1}^1  \frac{T_0(x + z c_* t)}{|c'(p_j)|}
 {\rm d} z, \qquad |c(p_j)| = c_*|z|.
\ee

We calculate the speed of heat front in one-dimensional chains.
Assume that function~$c(p)$ is limited. Therefore there exist a maximum value
of~$|x|$ such that the second  equation from formula~\eq{delt_1D} has a
solution. This value correspond to the heat front. From formulas~\eq{delt_1D}
it follows that the heat front of fundamental solution propagates with finite
speed~$c_{*}$, equal to the maximum group
velocity:
\be{}
 c_{*}  = \max_p{|c(p)|}.
\ee
The general solution corresponding to the initial temperature
distribution~$T_0(x)$ is given by formula~\eq{sol using delta 1D}.
Assume that~$T_0(x)$ is nonzero on the interval~$[x_{min}; x_{max}]$. Then from
formula~\eq{sol using delta 1D} it follows that at time~$t$ the temperature is
nonzero on the interval~$[x_{min}-c_* t; x_{max} + c_*t]$.  Therefore the heat
front in one-dimensional chains propagates with constant speed equal to the
maximum group velocity.\footnote{This result was also obtained in paper~\cite{Gendelman
2011 nonstationary substrate} using asymptotic analysis.}

\subsection{Two-dimensional scalar lattices}
In the present section, we derive the fundamental solution of the heat transfer
problem for
two-dimensional scalar lattices. The following distribution of initial
temperature is considered:
\be{IC delta}
  T_0 = A \delta(\rr) = A\delta(x)\delta(y),
\ee
where~$x$,$y$ are Cartesian coordinates.
Substitution of the initial conditions~\eq{IC delta} into
formula~\eq{sol slow}  yields
\be{sol pl HH}
  T_S = \frac{A}{16\pi^2} \int_{-\pi}^{\pi} \int_{-\pi}^{\pi}
  \Bigl(\delta\(\rr + \Vect{c} t\) + \delta\(\rr - \Vect{c} t\)\Bigr)
  {\rm d} p_1 {\rm d} p_2.
\ee
Radius-vector~$\rr$ and vector of group velocity~$\Vect{c}$ are represented as
\be{}
\rr = x\Vect{i} + y \Vect{j}, \qquad \Vect{c} = c_{x} \Vect{i} + c_{y}
\Vect{j},
\ee
where~$\Vect{i}$, $\Vect{j}$ are unit vectors corresponding to~$x$ and~$y$
axes. Then changing the integration variables~$(p_1, p_2) \rightarrow  (c_{x},
c_{y})$ in formula~\eq{sol pl HH}
and calculating the integral using the identity~\eq{ident delta}, we obtain
\be{sol pl HH1}
\begin{array}{l}
  \DS T_S = \frac{A}{16\pi^2 t^2} \sum_j \frac{1}{|G(p_1^j,p_2^j)|},
  \qquad G = \frac{\partial c_{x}}{\partial p_1} \frac{\partial c_{y}}{\partial
  p_2} - \frac{\partial c_{x}}{\partial p_2} \frac{\partial c_{y}}{\partial
  p_1},
  \\[4mm]
  \left[
  \begin{aligned}
  c_{x} &= \frac{x}{t},
  \quad
  c_{y} = \frac{y}{t},
  \\
  c_{x} &= -\frac{x}{t},
  \quad
  c_{y} = -\frac{y}{t},
\end{aligned}
\right.
  \end{array}
\ee
where~$G$ is the Jacobian of the transformation; square bracket stands for
logical ``or''; summation is carried out over the real roots~$p_1^j,p_2^j \in [-\pi;\pi]$ of the last two
equations.

Two facts follow from formulas~\eq{sol
pl HH1}. Firstly, the  temperature
is nonzero \edited{inside the circle}:
\be{ineq 2D}
 \(\frac{x}{c_* t}\)^2 + \(\frac{y}{c_* t}\)^2 \leq 1, \qquad c_*^2 =
 \max_{p_1,p_2} \(c_x^2 + c_y^2\).
\ee
Secondly, the temperature at the central point~$x=0, y=0$ decays as~$1/t^2$.

Thus the fundamental solution of the heat transport problem for two-dimensional
scalar lattices is given by formulas~\eq{sol pl HH1}. The solution has circular
front propagating with maximum group velocity~$c_*$. For example, the
fundamental solution for stretched square lattice performing out-of-plane
vibrations is obtained in section~\ref{sect fund square}.

\section{Example: length-dependence of the effective heat conductivity~(unsteady problem)}\label{sect
divergence heat conductivity}
From formula~\eq{sol slow} it follows that the heat transfer in
scalar lattices is ballistic and it can not be described by the Fourier law. In
this case, the notion of heat conductivity is ambiguous. Therefore the heat
conductivity can be defined differently in different problems. In papers~\cite{Lebowitz1967,Nakazawa1970,
Lee_Dhar_2005, Freitas Paz 2014} it is shown that in the steady-state the heat conductivity in harmonic
crystals is proportional to length of the system~(distance between heat
reservoirs). The main goal of the present section is to show that similar behavior of the heat conductivity is observed in {\it unsteady} problems.

Thermal contact of two half-spaces having different initial temperatures
is considered. Initial distribution of temperature has the form
\be{IC c/h}
   T_0(x) = T_1 + (T_2-T_1)H(x),
\ee
where~$H$ is the  Heaviside function; $T_1, T_2$ are initial temperatures of
the half-spaces~$x<0$ and $x>0$ respectively.
We define the heat conductivity as follows
\be{heat cond}
 \lambda = -\frac{\int_{-L}^{L} h(x,t) {\rm d} x}{T_S(L)- T_S(-L)},
\ee
where~$h$ is a projection of the heat flux on the~$x$-axis\footnote{The
relation between the heat flux, forces and particle velocities is not used in
the present derivations.};
$L$ is a half-length of averaging interval. If the Fourier law is valid, then
equation~\eq{heat cond} is satisfied identically and the heat conductivity is
independent on length~$L$. We show that for scalar lattices it is not the
case.

We calculate the heat flux~$h$ using continuum equation of energy
balance\footnote{Macroscopic mechanical deformation of the lattice and
volumetric heat sources in the present model are absent.}:
\be{dotT}
   \rho \dot{U} = -h', \qquad \rho = \frac{M}{V},
\ee
where~$V$ is a volume per particle.
The internal energy per unit mass~$U$ is calculated using formula\footnote{From
the virial theorem it follows that kinetic and potential energies per particle
are equal to~$k_B T/2$. Then the total energy per particle is equal to~$k_B
T$.}:
\be{U}
   U = \frac{k_B T_S}{M}.
\ee
Substituting formula~\eq{U} into equation of energy balance~\eq{dotT} yields
\be{hT}
  \frac{k_B}{V} \dot{T}_S= -h'.
\ee
Formula~\eq{hT} is used for calculation of the heat flux for given temperature
distribution.

In the case of one-dimensional initial temperature distribution~$T_0(x)$ in the
$d$-dimensional lattice, the general solution~\eq{sol slow} takes the form
\be{sol pl}
T_S = \frac{1}{4(2\pi)^d} \int_{-\pi}^{\pi}\Bigl(T_0\(x + c_{x} t\)
        + T_0\(x - c_{x} t\)\Bigr) {\rm d} p_1 ... {\rm d} p_d,
        \qquad
        c_x = \Vect{c}\cdot\Vect{i},
\ee
where~$\Vect{i}$ is a unit vector directed along~$x$-axis. Using
formula~\eq{sol pl} we show that the solution of the problem with initial
distribution
of temperature~\eq{IC c/h} is self-similar:
\be{}
\begin{array}{l}
  \DS T_S
  = \frac{T_1}{2} +
  \frac{T_2-T_1}{4(2\pi)^d} \int_{-\pi}^{\pi}
  \left[H\(\frac{x}{t}-c_{x}\) + H\(\frac{x}{t}+c_{x}\) \right] {\rm d} p_1 ...
  {\rm d} p_d
  =
  T_S\(\frac{x}{t}\).
  \end{array}
\ee
Integrating both parts of formula~\eq{dotT} from $-\infty$ to $x$ and assuming
that~$h(-\infty)=0$, we obtain:
\be{Hxt}
 h = -\frac{k_B}{V}\int_{-\infty}^x \frac{{\rm d}}{{\rm d} t}T_S\(\frac{z}{t}\)
 {\rm d} z = \frac{k_B}{V}\int_{-\infty}^{\frac{x}{t}} y T_S\(y\) {\rm d} y
 \quad
 \Rightarrow
 \quad
 h = h\(\frac{x}{t}\).
\ee
Here prime denotes the derivative with respect to~$x/t$. Formula~\eq{Hxt} shows
that the heat flux is also self-similar.

The heat conductivity is calculated using formula~\eq{heat cond}.
We choose~$L$ equal to the distance traveled by the heat front, i.e.~$L=c_*t$,
where $c_*$ is the maximum group
velocity\footnote{The fact that velocity of the heat front is equal to maximum group velocity, $c_*$, follows from fundamental solutions obtained above.}.
 Then~$T_S(L)- T_S(-L) = \frac{1}{2}\(T_2-T_1\)$. Substituting this expression into the definition
 of the heat conductivity~\eq{heat cond} and taking into
 account formula~\eq{Hxt}, yields:
\be{lambda vs l}
 \lambda = -\frac{2L}{T_2-T_1}\int_{-1}^{1} h\(z\) {\rm d} z \quad \Rightarrow
 \quad \lambda \sim L.
\ee
Formula~\eq{lambda vs l} shows that the effective heat conductivity linearly
diverges with length~$L$. \edited{Note that in anharmonic systems the dependence of effective heat conductivity on system size is nonlinear~(see e.g. papers~\cite{Lepri 2003, Li PRL 2003}).}

Thus we show that in the unsteady problem considered above the heat conductivity exhibits the same behavior as in steady problems considered in earlier works~\cite{Lebowitz1967, Nakazawa1970, Freitas Paz 2014}. Note that formula~\eq{lambda vs l} is derived for {\it any} scalar lattice described
by equations of motion~\eq{EM}.

\section{Example: one-dimensional chain}
Consider a one-dimensional chain with nearest-neighbor interactions. The equation of motion reads
\be{}
 \ddot{u}(x) = \omega_*^2 \Bigl( u(x + a) - 2 u(x) + u(x -a)\Bigr).
\ee
In this case operator~$\Ds$ is given by formula~\eq{om 1D}.

Short time behavior of kinetic temperature is described by integral~\eq{fast gen}.
Substituting parameters~\eq{om 1D}
into formula~\eq{fast gen} after integration, yields
\be{fast 1D}
  T_F = \frac{1}{2}T_0\(x\)J_0(4\omega_* t),
\ee
where $J_0$ is the Bessel function of the first kind. Formula~\eq{fast 1D}
shows that~$T_F$ asymptotically tends to zero inversely proportional to the square root of time\footnote{This
fact follows from the asymptotic representation of Bessel function~$J_0$.}.
This result has originally been obtained in paper~\cite{Krivtsov
2014 DAN}.

Ballistic heat transfer is described by formula~\eq{sol slow}.
Substitution of formulas~\eq{om 1D}
into expression for group velocity~\eq{group vel general}, yields
\be{group vel 1D chain}
    c =\omega_* a  \cos \frac{p}{2} {\rm sign} p. 
\ee
Then the general solution has form
\be{sol 1D gen AK}
    T_S = \frac{1}{2\pi} \int_0^{\frac{\pi}{2}} \Bigl( T_0(x + c_*t \cos p) +
    T_0(x - c_*t \cos p)\Bigr) {\rm d} p, \qquad c_* =\omega_* a.
\ee
Corresponding fundamental solution is obtained using formula~\eq{delt_1D}:
\be{delt_1D Kr}
    T_S = \frac{A}{2\pi c_* t \sqrt{1-\(\frac{x}{c_*t}\)^2}}.
\ee
Formulas~\eq{sol 1D gen AK}, \eq{delt_1D Kr} coincide with results obtained in
paper~\cite{Krivtsov DAN 2015}.

Thus in the case of the one-dimensional chain with nearest-neighbor interactions, we reproduce the results obtained in papers~\cite{Krivtsov
2014 DAN, Krivtsov DAN 2015}.

\edited{{\bf Remark.}
The difference between time scales corresponding to fast and slow processes is clearly demonstrated
using the following example. Consider initial conditions~\eq{IC kappa} corresponding
to sinusoidal distribution of temperature:
\be{T sin 1D}
    T_0(x) = B_0\sin\frac{2\pi x}{L} + B_1,
\ee
where $L$ is wave-length of initial temperature distribution; $B_1 \geq B_0$.
Substitution of the initial conditions~\eq{T sin 1D} into formulas~\eq{fast 1D}, \eq{sol 1D gen AK} after algebraic transformations
yields:
\be{T sin 1D sol}
 T =
 \frac{B_1}{2}\bigl(1 + J_0(4\omega_* t)\bigr)
 +
 \frac{B_0}{2} \(J_0(4\omega_* t) + J_0\(\frac{2\pi c_* t}{L}\)\)\sin\frac{2\pi x}{L}.
\ee
Formula~\eq{T sin 1D sol} contains two dimensionless times~(time scales) --- $\omega_* t$ and $c_*t/L$. The first time scale is determined by frequencies of vibrations of individual atoms. The second time scale is determined by a time required for a wave to travel distance~$L$.
The ratio of these time scales, being proportional to~$L/a$, is a large parameter. Therefore time scales of fast and slow thermal processes are well separated.}

\section{Example: out-of-plane vibrations of a square lattice}\label{sect
square}
\subsection{General formulas}
In the present section, we consider out-of-plane vibrations of a stretched
square lattice.
Initial radius-vectors of the particles have the form:
\be{rad vect}
 \xx_{n,m} = a\(n\Vect{i} + m\Vect{j}\),
\ee
where~$\Vect{i}$, $\Vect{j}$ are orthogonal unit vectors; $a$ is an initial
distance between the nearest neighbors. Particles are connected to their
nearest neighbors by linear springs. Equilibrium length of the springs is less
than~$a$, i.e. the lattice is stretched\footnote{Otherwise the out-of-plane
vibrations of the lattice are nonlinear.}. Then linearized equations for
 out-of-plane vibrations of the lattice have the
 form\footnote{In harmonic approximation, in-plane and out-of-plane vibrations
 of the lattice are independent.
 The in-plane vibrations of the lattice are beyond the scope of the present
 paper.}:
 \be{EM sq}
  \ddot{u}_{n,m} =\Ds {u}_{n,m}, \qquad  \Ds {u}_{n,m} = \omega_*^2 \(u_{n+1,
  m} + u_{n, m+1} - 4u_{n,m} + u_{n-1,m} + u_{n,m-1}\),
 \ee
where~$u_{n,m} = u(\xx_{n,m})$ is a component of displacement normal to
the lattice plane.
It is seen that equation~\eq{EM sq} is a particular case of equation~\eq{EM},
where parameters~$\omega_*$, $\va$, $b_{\alpha}$ are determined by
formula~\eq{ba square}.

The dispersion relation and group velocity are calculated using
formulas~\eq{ba square}, \eq{disp rel},  \eq{group vel general}:
\be{group vel square}
\omega = 2 \omega_* \sqrt{\sin^2\frac{p_1}{2} + \sin^2\frac{p_2}{2}},
\quad
 \Vect{c} = \frac{\DS c_*\(\sin p_1 \Vect{i}  +\sin p_2 \Vect{j}\)}{\DS
 2\sqrt{\sin^2\frac{p_1}{2} + \sin^2\frac{p_2}{2}}},
 \quad
 \kk = \frac{1}{a}\(p_1 \Vect{i} + p_2 \Vect{j}\),
\ee
where~$c_* = \omega_* a$ is the maximum group velocity.

\subsection{Short time behavior of kinetic temperature~(fast process)}\label{fastsq}
In the present section, we consider short time behavior of kinetic temperature in the uniformly heated lattice.

The initial conditions
for particles corresponding to uniform distribution of instantaneous
temperature~$T_0$ read
\be{IC part uniform}
   u_{n,m} = 0, \qquad  v_{n,m} = v_0,
\ee
where $v_0$ is a random quantity with dispersion~$\av{v_0^2} = k_B T_0/M$.
In this case initial kinetic and potential energies of the system are not
equal. Equilibration of energies leads to oscillations of the kinetic
temperature  described by formula~\eq{fast gen}. Substitution
of dispersion relation~\eq{group vel square}
into formulas~\eq{fast gen}, \eq{sol slow} yields:
\be{fast sq}
     T = T_S + T_F,
     \qquad
     T_S = \frac{T_0}{2},
     \qquad
     T_F =  \frac{T_0}{2\pi^2} \int_{0}^\pi\int_{0}^\pi
     \cos\(4\omega_* t \sqrt{\sin^2\frac{p_1}{2} + \sin^2\frac{p_2}{2}}\)
    {\rm d}p_1{\rm d}p_2.
\ee
Note that formula~\eq{fast sq} {\it exactly} describe the behavior of kinetic temperature in the uniformly heated lattice.

\begin{figure*}[htb]
\begin{center}
\includegraphics*[scale=0.4]{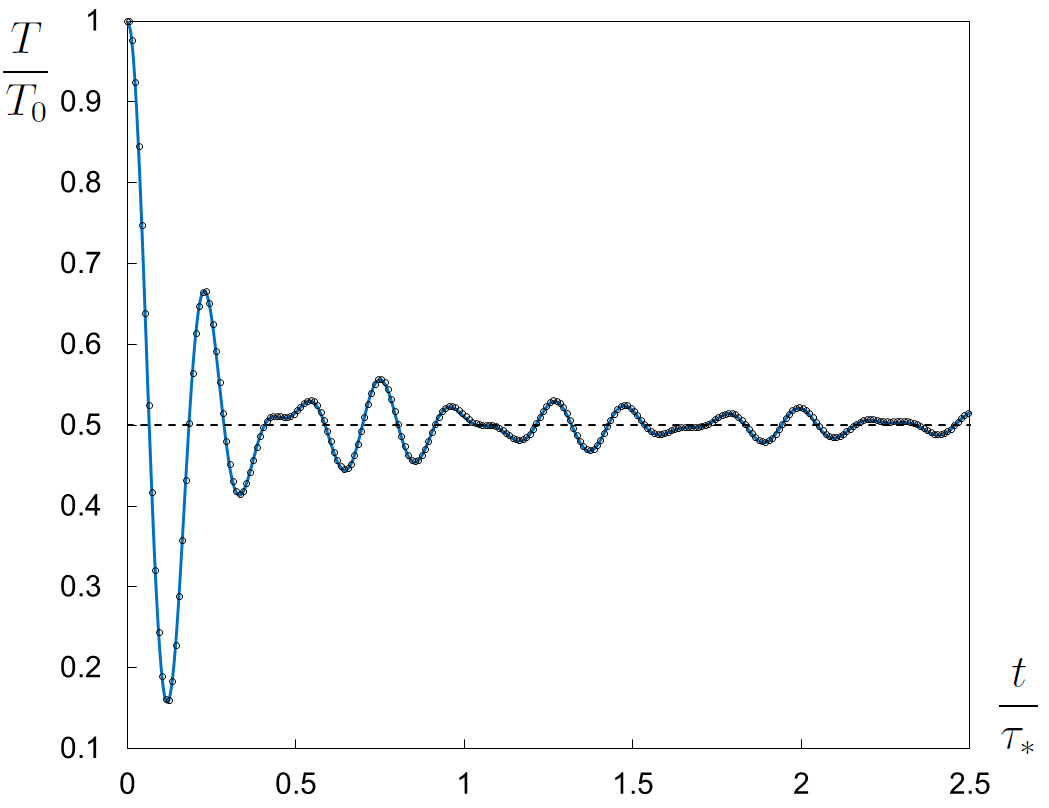}
\caption{Short time behavior of kinetic temperature in the uniformly heated stretched square
lattice.
Solid line --- analytical solution~\eq{fast sq}, dashed line --- $T_S = T_0/2$,  circles ---  numerical solution of
lattice dynamics equations~\eq{EM sq}.}
\label{fast_square_latt}
\end{center}
\end{figure*}

In order to check formula~\eq{fast sq}, we compare the results
with numerical solution of equations of motion~\eq{EM sq}. Leap-frog
integration scheme with time-step equal to~$0.005 \tau_*$,
$\tau_*=2\pi/\omega_*$ is used. Periodic boundary conditions in both directions
are applied.  Square periodic cell contains~$10^6$ particles. During the
simulation the kinetic temperature of the entire system is calculated. The
dependence of temperature on time is shown
in Figure~\ref{fast_square_latt}. Every point on the plot corresponds to
average
over~$10$ realizations with different initial conditions. The standard error of
the mean
is of order of circle diameter.
Figure~\ref{fast_square_latt} shows that analytical solution~\eq{fast sq}
coincides with results of numerical solution of lattice dynamics
equations~\eq{EM sq}.

Temperature oscillations caused by equilibration of kinetic and potential
energies decays in time. Characteristic time of the decay is of order of
several periods~$\tau_*$. Multiplying the temperature by time it can be shown
that deviation from the stationary value decays as~$1/t$. The same process in
one-dimensional chain decays as~$1/\sqrt{t}$. Heat propagation is a much slower
process. For example, during~$\tau_*$ the heat front passes the distance equal
to~$2\pi a$, which is small from macroscopic point of view.

Thus the example considered in the present section shows that oscillations of
temperature and ballistic heat transfer have different time scales.  Therefore the notions ``fast process'' and  ``slow process'' are used. Comparison with results of computer simulations show that equation~\eq{fast sq} accurately describe the fast process.

\subsection{Ballistic heat transfer~(slow process)}
In the present section, we investigate the ballistic heat transfer described by formula~\eq{sol slow} in the stretched square lattice.

\subsubsection{Fundamental solution of the planar problem}\label{sect delt}
\begin{figure*}[htb]
\begin{center}
\includegraphics*[scale=0.4]{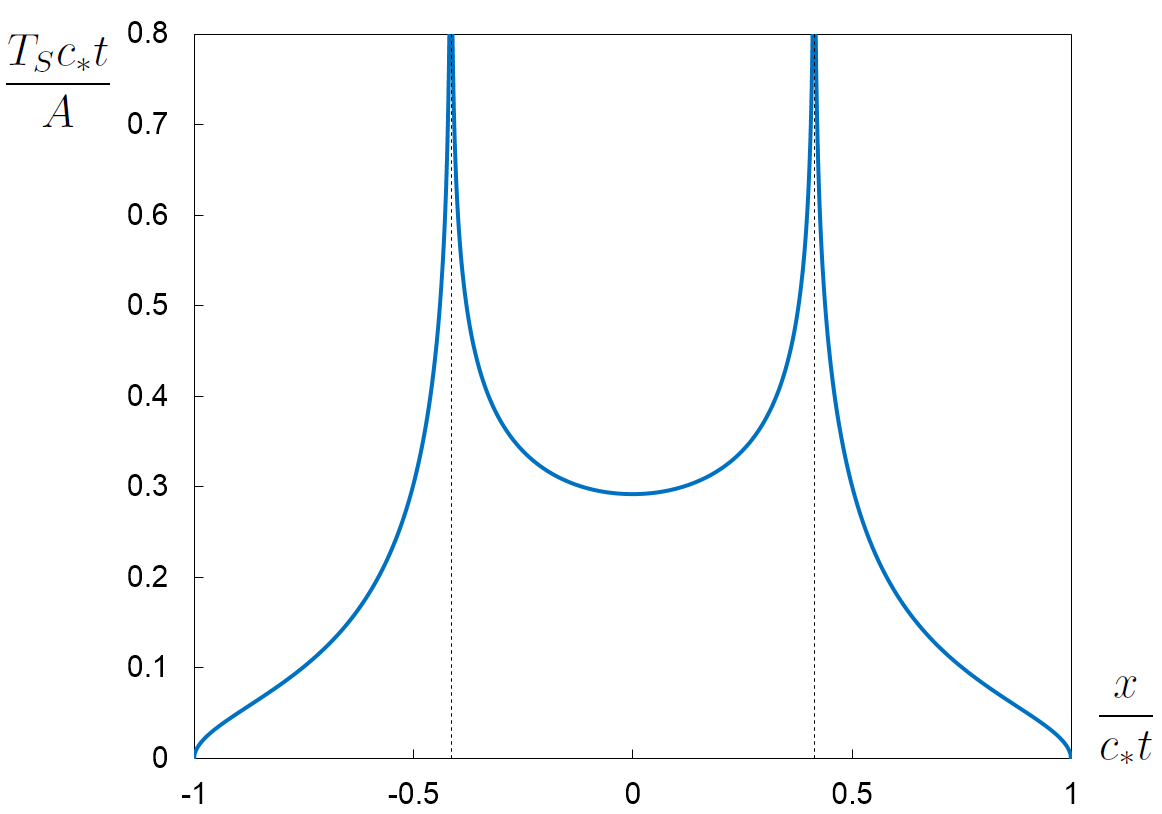}
\caption{Solution of the heat transfer problem with initial conditions~\eq{IC
delta plane}. Solid line --- formula~\eq{an sol delta}; dashed lines---
vertical asymptotes at~$|\tilde{x}|=\sqrt{2}-1$.}
\label{delta_square_latt}
\end{center}
\end{figure*}

We derive the fundamental solution of planar heat
transport problem for the stretched square lattice.
The following initial temperature distribution is considered:
\be{IC delta plane}
     T_0(x) = A\delta(x),
\ee
where $x$ is directed along the basis vector~$\Vect{a}_1= a\Vect{i}$.
Substituting the initial conditions into formula~\eq{sol pl} and taking into
account formula~\eq{group vel square},
yields:
\be{sol delta gen}
T_S = \frac{A}{4\pi^2} \int_0^{\pi} \int_0^{\pi}
 \Bigl(\delta\(x-c_{x}t\) + \delta\(x+c_{x}t\)\Bigr){\rm d}p_1{\rm d}p_2,
 \qquad
    c_x = \frac{\DS c_* \sin p_1}{\DS 2\sqrt{\sin^2\frac{p_1}{2} +
    \sin^2\frac{p_2}{2}}}.
\ee
We make a substitution~$\beta = \sin^2\frac{p_1}{2}$, $\gamma =
\sin^2\frac{p_2}{2}$ and consider the case~$t>0$, $x>0$. Solution for~$x<0$ is
obtained using symmetry of the problem.
Then~\eq{sol delta gen} takes the form:
\be{delt sq bg}
T_S =
 \frac{A}{4\pi^2}\int_{0}^{1}\int_0^{1}  \frac{
 \delta\(\tilde{x}-\sqrt{\frac{\beta(1-\beta)}{\beta + \gamma}}\)}
  {\sqrt{\beta\gamma(1-\beta)(1-\gamma)}}{\rm d}\beta{\rm d}\gamma, \qquad
  \tilde{x} = \frac{x}{c_*t}.
\ee
One of the integrals is evaluated using the identity~\eq{ident delta}. The
argument of delta-function in formula~\eq{delt sq bg} has roots given by the
following equation
\be{gam}
 \gamma =\frac{\beta}{\tilde{x}^2}  \(1-\tilde{x}^2-\beta\).
\ee
By the definition~$0 \leq \gamma \leq  1$. Then formula~\eq{gam} yields the
inequalities for~$\beta$:
\be{ineq beta}
 \beta \leq 1 - \tilde{x}^2,
  \qquad
 \beta^2 - (1-\tilde{x}^2)\beta + \tilde{x}^2 \geq 0.
\ee
Solving the inequality~\eq{ineq beta} and using the identity~\eq{ident delta},
we obtain:
\be{an sol delta}
\begin{array}{l}
\DS T_S =
\frac{A}{2\pi^2 c_{*} t|\tilde{x}|}\begin{cases}
\DS
  f(0, 1-\tilde{x}^2), \quad \sqrt{2}-1 \leq |\tilde{x}| \leq 1,
\\[4mm]
\DS f(0, \beta_1) + f(\beta_2,1-\tilde{x}^2), \quad |\tilde{x}| \leq
\sqrt{2}-1, \\[4mm]
\end{cases}
\\[4mm]
\DS
 \beta_{1,2} = \frac{1}{2}\(1-\tilde{x}^2 \mp \sqrt{(1-\tilde{x}^2)^2 -
 4\tilde{x}^2}\),
 \\[4mm]
\DS f(\xi_1,\xi_2) = \int_{\xi_1}^{\xi_2}
\(\frac{1-\beta}{(1-\tilde{x}^2-\beta)(\tilde{x}^2
-\beta(1-\tilde{x}^2-\beta))}\)^{\frac{1}{2}} {\rm d} \beta,
\end{array}
\ee
and~$T_S=0$ for~$|\tilde{x}| \geq 1$.
Formula~\eq{an sol delta} shows that function~$T_S c_{*} t / A$ depends only on
the self-similar variable~$\tilde{x}$~(see
figure~\ref{delta_square_latt}).
It is seen from figure~\ref{delta_square_latt} that the heat front moves with
constant speed equal to~$c_{*}$. Temperature have singularities at the
points~$|\tilde{x}| = \sqrt{2}-1$.
Note that in one-dimensional chain the temperature in similar problem has
singularities
at the heat front~$|\tilde{x}|=1$~(see formula~\eq{delt_1D Kr}).

\subsubsection{Thermal contact of hot and cold half-planes} \label{sect
Heaviside}
Consider thermal contact of two half-planes with
initial temperatures~$T_1$ and
$T_2$~(see formula~\eq{IC c/h}). This problem is important, because it is closely related to classical definition of temperature~\cite{Hoover_stat_phys}. By the definition, temperatures of two bodies in thermodynamics equilibrium are equal. The problem considered below demonstrates the transition to thermodynamic equilibrium.

%
%
Substituting initial conditions~\eq{IC c/h} into the solution~\eq{sol pl} and
taking into account properties of the Heaviside function
and function~$c_{x}$, yields:
\be{sol H gen 1}
T_S = \frac{1}{4}\(T_1 + T_2\) +   \frac{1}{2}\(T_2-T_1\)w\(\frac{|x|}{t}\) {\rm sign}\(x\),
\qquad
w = \frac{1}{2\pi^2} \int_0^{\pi}\int_0^{\pi} H\(|x|-c_{x}t\){\rm d}p_1{\rm d}
p_2.
\ee
We make the substitution~$\beta = \sin^2\frac{p_1}{2}$, $\gamma =
\sin^2\frac{p_2}{2}$, then
\be{T1}
w = \frac{1}{2\pi^2}\int_{0}^{1}\int_0^{1}  \frac{
H\(|\tilde{x}|-\sqrt{\frac{\beta(1-\beta)}{\beta + \gamma}}\)}
  {\sqrt{\beta\gamma(1-\beta)(1-\gamma)}}{\rm d}\beta{\rm d}\gamma.
\ee
Integrand in formula~\eq{T1} is nonzero if the following inequality is
satisfied:
\be{ineq1}
 \gamma \geq \frac{\beta}{\tilde{x}^2}  \(1-\tilde{x}^2-\beta\).
\ee
The inequality~\eq{ineq1} is satisfied identically for~$\beta > 1-\tilde{x}^2$;
$\beta$ also satisfies the second inequality from~\eq{ineq beta}. Then
evaluation of the integral with respect
to~$\beta$, yields:
\be{sol Hev}
\begin{array}{l}
\DS w =  \frac{1}{4} + \frac{1}{2\pi}\arcsin|\tilde{x}| -
\begin{cases}
   g(0, 1-\tilde{x}^2),~~\sqrt{2}-1 \leq |\tilde{x}| \leq 1,
\\[4mm]
\DS  \frac{\arcsin\beta_2 - \arcsin\beta_1}{2\pi} + g(0, \beta_1)
   + g(\beta_2, 1-\tilde{x}^2),~~|\tilde{x}| \leq \sqrt{2}-1, \\[4mm]
\end{cases}
\\[4mm]
\DS g(z_1,z_2) =
\frac{1}{2\pi^2}\int_{z_1}^{z_2}\frac{\arcsin\(\frac{2\beta}{\tilde{x}^2}
\(1-\tilde{x}^2-\beta\) -1\)}{\sqrt{\beta(1-\beta)}} {\rm d} \beta,
\end{array}
\ee
where~$\beta_1, \beta_2$ are defined by formula~\eq{an sol delta}; $w =
\frac{1}{2}$ for~$|\tilde{x}| \geq 1$.

Thus solution of the problem is given by formulas~\eq{sol H gen 1},
\eq{sol Hev}. It is seen that the solution~\eq{sol Hev} is self-similar and
it depends on~$\tilde{x} = x/(c_*t)$.

\begin{figure*}[htb]
\begin{center}
\includegraphics*[scale=0.45]{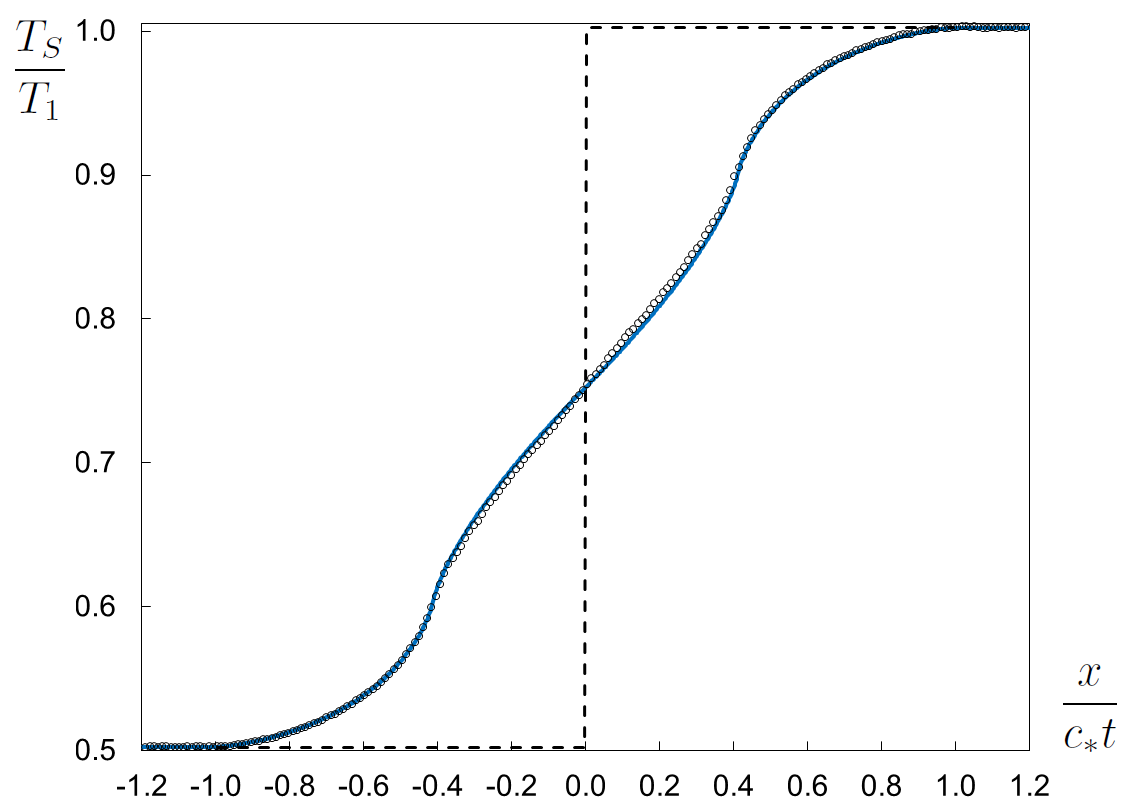}
\caption{Contact of hot and cold half-planes: self-similar temperature profile.
Line --- analytical solution~\eq{sol Hev}, circles --- numerical solution of
lattice dynamics equations~\eq{EM sq}.}
\label{Hs_square_latt}
\end{center}
\end{figure*}

We check the accuracy of formulas~\eq{sol H gen 1}, \eq{sol Hev} using
numerical
solution of lattice dynamics equations~\eq{EM sq}. Without loss of generality
we put~$T_2=2T_1$.
In this case, initial conditions for particles has the form:
\be{IC part Heav}
\begin{array}{l}
  u_{n,m} = 0,
  \qquad
  v_{n,m} =
  \begin{cases}
   v_0, \quad n < 0,
\\[4mm]
  \sqrt{2} v_0  \quad n >= 0.
  \end{cases}
  \end{array}
\ee
where $v_0$ is a random quantity with dispersion~$\av{v_0^2} = k_B T_1/M$.
Periodic boundary conditions are used. The periodic cell contains~$4\cdot10^6$
particles~($4\cdot10^2$ in the $x$-direction and $10^4$ in the $y$-direction).
In order to compute temperature, we consider~$10^3$ realizations with initial
conditions~\eq{IC part Heav}. Then temperature is computed by formula~\eq{kin
temp}, where mathematical expectation is approximated by an average over
realizations. Since the solution is self-similar, then it is sufficient to
consider only one moment of time. Temperature distribution at~$t=15\tau_*$ is
computed. At this moment oscillations of kinetic
temperature described in section~\ref{fastsq} practically vanish. The
temperature distribution is additionally averaged in $y$ direction. Comparison
of numerical results with analytical solution~\eq{sol Hev} is
shown in figure~\ref{Hs_square_latt}.
Small differences between analytical and numerical solutions are observed in
the vicinity of the central point~$x=0$. At this point the temperature has
large gradient~(initially it is infinite) and therefore the long-wave
approximation looses the accuracy. Far from the central point, analytical
solution~\eq{sol Hev} almost coincide with numerical results.

\subsubsection{Rectangular distribution of initial temperature. Thermal waves}
\begin{figure*}[htb]
\begin{center}
\includegraphics*[scale=0.45]{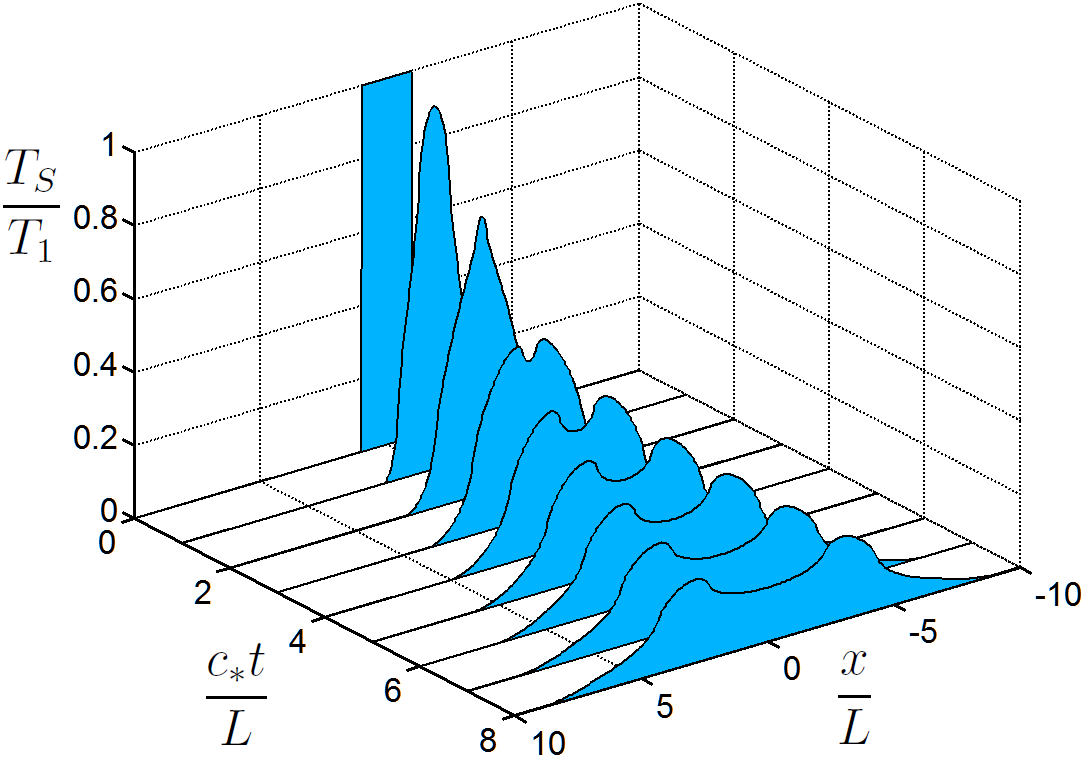}
\caption{Evolution of rectangular initial temperature distribution~\eq{pulse} in scalar
square lattice.}
\label{squarepulse}
\end{center}
\end{figure*}

In the present section, we demonstrate once again that the heat transfer in stretched  square lattice is
ballistic. Rectangular distribution of initial temperature is considered:
\be{pulse}
     T_0(x) = 2T_1 \Bigl(H(x+L) - H(x-L)\Bigr),
\ee
where~$L$ is a half-length of the interval with nonzero initial temperature.
Solution of the problem with initial conditions~\eq{pulse} is obtained
using formula~\eq{sol Hev} and the superposition principle. The resulting
distribution of temperature at several moments of time
is shown in figure~\ref{squarepulse}.
Figure~\ref{squarepulse} clearly shows two ``thermal waves'' traveling in
opposite directions. Peaks of the temperature distribution move with constant speed equal to~$(\sqrt{2}-1)c_*$. This fact can be used for validation of presented theory in future laboratory experiments.


\subsubsection{Sinusoidal distribution of initial temperature}\label{sect_sinus}
In the present section, we consider the decay of initial sinusoidal temperature distribution. This problem is important, because it allows to clearly demonstrate that the diffusive and hyperbolic~\cite{CV1, CV2} heat transfer equations are not applicable to harmonic crystals. Similar problem for harmonic one-dimensional chains is considered in papers~\cite{Krivtsov DAN 2015, Krivtsov arxive 2015, Gendelman 2011 nonstationary substrate}. \edited{We also demonstrate the difference between time scales of fast and slow thermal processes.}

Consider the following distribution of initial temperature:
\be{T sin}
    T_0(x) = B_0\sin\frac{2\pi x}{L} + B_1,
\ee
where $L$ is wave-length of initial temperature distribution; $B_1 \geq B_0$.
Fourier's law as well as the hyperbolic heat transfer equation~\cite{CV1},
\cite{CV2} predict that amplitude of sin decays exponentially.
In this section, we show using analytical solution~\eq{sol pl} and numerical
simulations that the amplitude decays inversely proportional to time.

Substituting the initial temperature distribution~\eq{T sin}
into the general solution~\eq{sol pl}, yields\footnote{The identity~$\sin(x\pm
y) = \sin x \cos y \pm \sin y\cos x$
is used for derivation.}:
\edited{
\be{sin analyt}
\begin{array}{l}
\DS T = B(t)\sin\frac{2\pi x}{L} + B_1,
\qquad
B(t) = B_F + B_S,
\\[4mm]
\DS B_F = \frac{B_0}{2\pi^2} \int_{0}^\pi\int_{0}^\pi
     \cos\(4\omega_* t \sqrt{\sin^2\frac{p_1}{2} + \sin^2\frac{p_2}{2}}\)
    {\rm d}p_1{\rm d}p_2,
\\[4mm]
\DS B_S =
\frac{B_0}{2\pi^2} \int_0^\pi\int_0^\pi
\cos\(\frac{\pi c_{*} t\sin p_1}{L \sqrt{\sin^2\frac{p_1}{2} +
\sin^2\frac{p_2}{2}}}\) {\rm d} p_1{\rm d}p_2.
\end{array}
\ee}
\edited{Formula~\eq{sin analyt} contains two dimensionless times~$\omega_* t$ and $c_*t/L$ corresponding to
fast and slow thermal processes.}

\begin{figure*}[htb]
\begin{center}
\includegraphics*[scale=0.45]{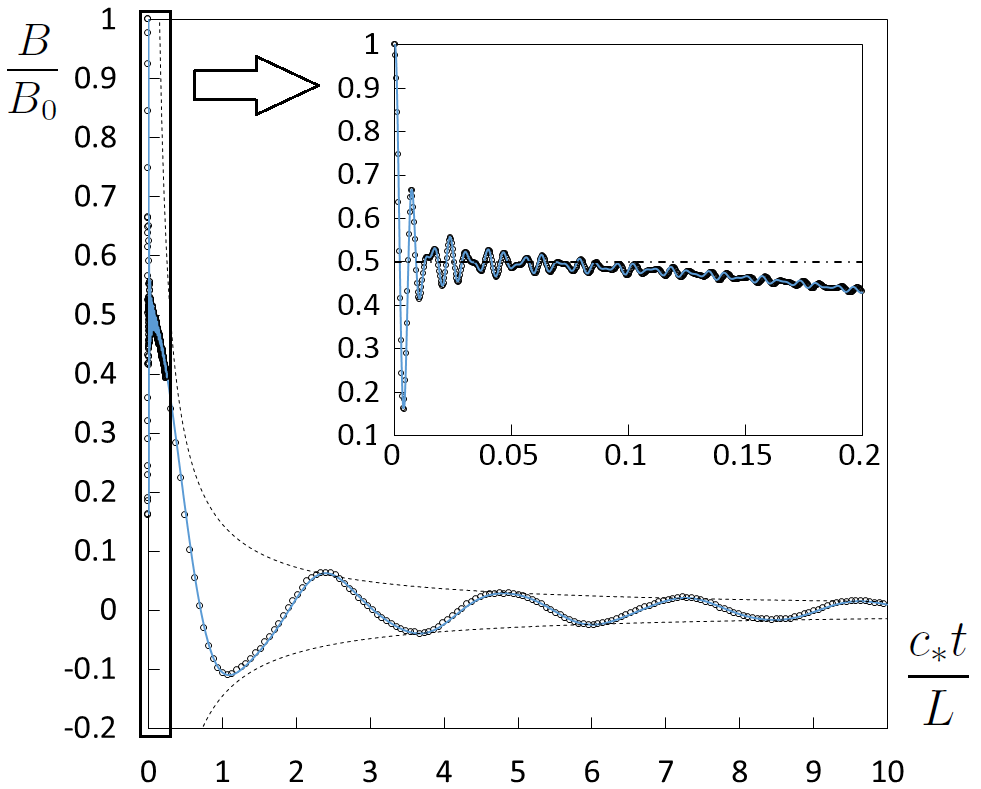}
\caption{Decay of amplitude of initial sinusoidal temperature distribution~\eq{T sin}. \edited{Short-time behavior of the amplitude is shown in the  subplot.}
Solid line --- analytical solution~\eq{sin analyt}, circles  --- numerical
solution of lattice dynamics equations~\eq{EM sq}, dashed lines ---
envelope~$\pm 0.145 L/(c_*t)$.}
\label{decayofsin}
\end{center}
\end{figure*}

We check the accuracy of formula~\eq{sin analyt} using numerical solution of
lattice dynamics equations~\eq{EM sq}. Particles have random initial velocities
corresponding to initial
temperature distribution~\eq{T sin}. Initial particle displacements are equal to zero. Periodic boundary conditions in both
directions are used. The periodic cell contains~$2\cdot10^6$ particles. The size of
the periodic cell in the $x$-direction is equal to~$L = 2\cdot10^2 a$. Results are
averaged over~$10^3$ realizations with different initial conditions.
The dependence of amplitude~$B$ on dimensionless time~$c_* t/L$ is shown in
figure~\ref{decayofsin}.
Every circle on the plot corresponds to average over realizations. Standard
error of \edited{the mean is of order of the diameter of the circle}.
Figure~\ref{decayofsin} shows that analytical solution~\eq{sin analyt}
practically coincides with results of numerical solution of lattice dynamics
equations~\eq{EM sq} \edited{at both short and large time scales}.

The amplitude of sinusoidal distribution of initial temperature in scalar
square lattice decays inversely proportional to time.
In the one-dimensional chain with nearest neighbor interactions, the same
oscillations are described by the Bessel function of the first
kind~\cite{Krivtsov DAN 2015, Krivtsov arxive 2015}, which decays inversely
proportional to square root of time. Therefore in both cases the amplitude
decays according to a power law,  while diffusive and hyperbolic~\cite{CV1, CV2} heat transfer equations predict exponential decay.

This example also shows that the heat transfer process
in scalar lattices is {\it irreversible}. At the same time, the general
solution of the heat transfer problem~\eq{sol slow} is symmetric with respect
to time, i.e. invariant with respect to substitution~$t$ by $-t$. This fact may serve for better understanding of the Loschmidt's~(reversibility)
paradox~\cite{Hoover chaos}.

\subsubsection{Fundamental solution}\label{sect fund square}
The fundamental solution of heat transport problem for two-dimensional scalar
lattices is given by formula~\eq{sol pl HH1}. In order to obtain the solution
for  the square lattice, we calculate the Jacobian~$G$ using formulas~\eq{sol pl
HH1}
and~\eq{group vel square}:
\be{J_isq}
 G = - \frac{c_*^2\(\cos p_1 \sin^4\frac{p_2}{2} + \cos p_2
 \sin^4\frac{p_1}{2}\)}{4\(\sin^2\frac{p_1}{2} + \sin^2\frac{p_2}{2}\)^2}.
\ee
We make two consecutive substitutions in formulas~\eq{sol pl HH1},
\eq{J_isq}:~$s_1=\sin^2\frac{p_1}{2}, s_2 = \sin^2\frac{p_2}{2}$ and $w =
s_1s_2$,
$q = s_1+s_2$. Then excluding~$w$ we obtain:
\be{T IC delt2}
\begin{array}{l}
\DS T_S = \frac{A}{2(\pi c_* t)^2} \sum_j \frac{q_j}{|q_j^2-\tilde{r}^2(q_j+1)|},
\quad
\\[5mm]
  \DS q_j^3 - 2\(\tilde{r}^2+1\)q_j^2 +  \bigl((\tilde{r}^2+1)^2 - 4
  \tilde{x}^2\tilde{y}^2 + 1\bigr)q_j - 2\tilde{r}^2 = 0,
 \end{array}
\ee
where $\tilde{r}^2 = 1-\tilde{x}^2-\tilde{y}^2$,
  $\tilde{x} = \frac{x}{c_* t}$, $\tilde{y} = \frac{y}{c_* t}$. Summation is
  carried out with respect to all real roots~$q_j$ of the given cubic equation
  such that~$q_j \in [0;2]$. Formula~\eq{T IC delt2} shows that the function~$T
  c_*^2 t^2/A$ is self-similar.

\begin{figure*}[htb]
\begin{center}
\includegraphics*[scale=0.45]{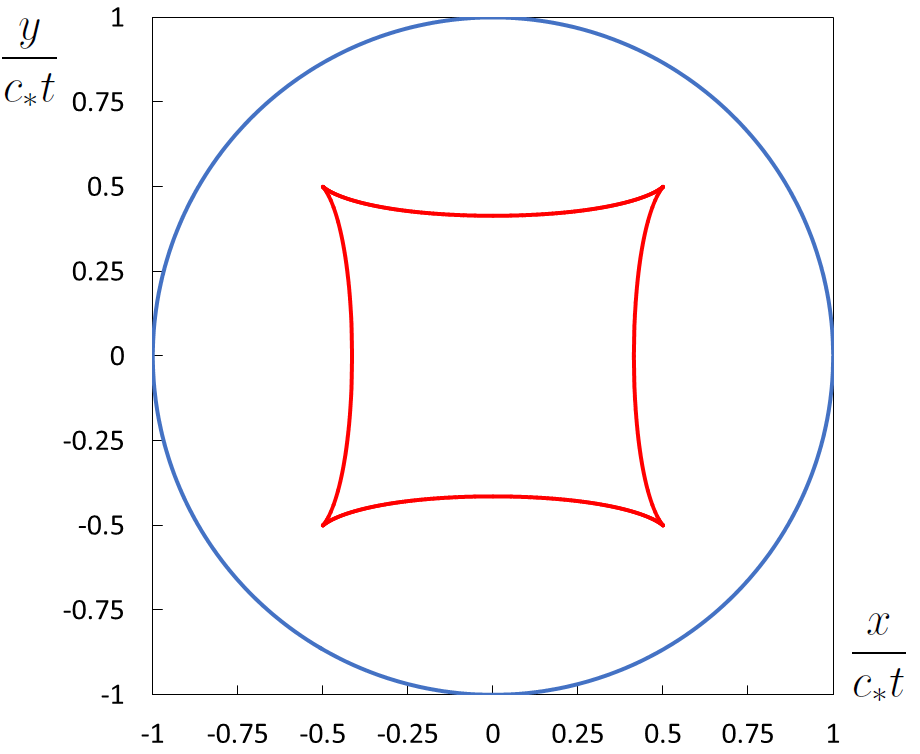}
\caption{Heat front~(circle) and line with infinite
temperature~(equation~\eq{deltinf})
corresponding to the fundamental solution~\eq{T IC delt2}.}
\label{2D delta inf}
\end{center}
\end{figure*}

 Formula~\eq{T IC delt2} gives closed-form fundamental solution for harmonic
 scalar square lattice. According to formula~\eq{T IC delt2}, the temperature
 is nonzero inside the circle~$\tilde{x}^2 +  \tilde{y}^2 \leq 1$.
It has singularity along the line determined by the following system of
equations:
\be{deltinf}
  \tilde{x}^2 + \tilde{y}^2 = 1- \frac{q^2}{q+1},
  \qquad
  \tilde{x}^2\tilde{y}^2  = \frac{2-q^2}{4(q+1)^2}.
\ee
The line~\eq{deltinf} is shown in figure~\ref{2D delta inf}. It
intersects~$\tilde{x}$-axis at the points~$\pm (\sqrt{2}-1)$.

\begin{figure*}[htb]
\begin{center}
\includegraphics*[scale=0.4]{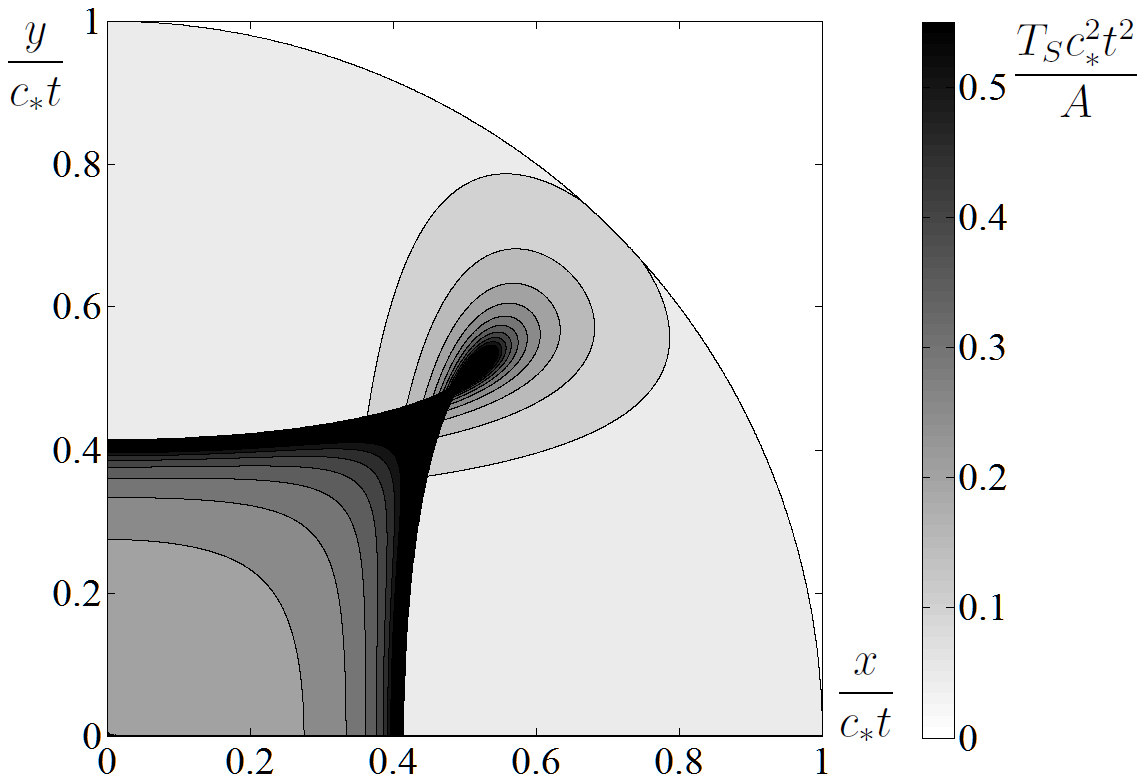}
\caption{Fundamental solution~\eq{T IC delt2} of the unsteady heat transfer
problem for scalar square lattice.}
\label{2D delta square latt}
\end{center}
\end{figure*}

Fundamental solution~\eq{T IC delt2} is symmetrical with respect to
axes~$\tilde{x}$ and $\tilde{y}$. Solution for positive~$\tilde{x}$,
$\tilde{y}$
is shown in figure~\ref{2D delta square latt}.
We check the accuracy of fundamental solution~\eq{T IC delt2} as follows.
Problems described in sections~\ref{sect delt}, \ref{sect Heaviside} are solved
using the convolution of the fundamental solution with corresponding initial
conditions. It is shown that the resulting temperature distribution coincides
with results obtained in sections~\ref{sect delt}, \ref{sect Heaviside}.

Thus the closed-form fundamental solution of unsteady heat transfer problem for
scalar square lattice is given by formula~\eq{T IC delt2}. We note the analogy
between our result~\eq{T IC delt2} and results obtained in papers~\cite{Melike, Harris 2008 SIAM}.
In papers~\cite{Melike, Harris 2008 SIAM},  spatial distribution of energy corresponding to the
fundamental solution of equations of motion~\eq{EM} is obtained using the Wigner
transform. The energy distribution is similar to temperature distribution shown
in figure~\ref{2D delta square latt}. Therefore there is an analogy between
deterministic lattice dynamics problem~\cite{Melike} and the unsteady heat
transfer problem discussed above.

\section{Conclusions}
An equation exactly describing the evolution of temperature field in {\it any} scalar lattice was derived. Using this equation, we have shown that the temperature field in a lattice with random initial velocities and zero initial displacements is represented as a sum of two terms~(see formulas~\eq{sol fast slow}, \eq{fast gen}, \eq{sol slow}).

The first term describes short time behavior of temperature. At short times, temperature performs decaying oscillations caused by redistribution of energy among kinetic and potential forms. These oscillations in different spatial point are independent. Characteristic time of decay is of order of ten periods of atomic vibrations.

The second term describes large time behavior of kinetic temperature associated with unsteady ballistic heat transfer. At large times, the temperature field  is represented as a
superposition of waves having a shape of initial temperature distribution and
traveling with the group velocity. The heat front propagates with constant speed
equal to the maximum group velocity. These observations are consistent with
results obtained in papers~\cite{Melike, Lukkarinen Spohn 2007, Dudnikova Spohn
transport eq, Harris 2008 SIAM, Lukkarinen 2016} by completely different means. Closed-form
fundamental solutions of the unsteady heat transfer problem for one- and
two-dimensional scalar lattices were derived.

The expression for the temperature field  has the same property as the
equations of motions: it is invariant to the substitution~$t \rightarrow -t$.
However thermal processes in scalar lattices are irreversible.
In order to illustrate this fact, an analytical solution for problem with
sinusoidal distributions of initial temperature in scalar square lattice was
derived. The solution shows that the amplitude of sinusoidal distribution
decays inversely proportional to time. Therefore the process is irreversible,
while it is described by the equation invariant to the substitution~$t
\rightarrow -t$.

Comparison of analytical results with numerical simulations  shows that presented theory describes the behavior of temperature field at both short and large time scales with high accuracy.

\section{Acknowledgements}

The authors are deeply grateful to M.B. Babenkov,  W.G. Hoover, S.N. Gavrilov, E.A. Ivanova, D.A. Indeytsev, M.L. Kachanov, O.S. Loboda, G.S. Mishuris, N.F. Morozov, and A. Politi for
useful discussions. \edited{Comments of the reviewers are highly appreciated.}

Numerical simulations have been carried out using
facilities of the Supercomputer Center ``Polytechnic'' at Peter the Great Saint Petersburg Polytechnic University.

This work was supported by the Russian Science Foundation (RSCF grant No. 17-71-10213).

\appendix

\section{Appendix. Equations for covariances}\label{eqcovariances}
In the present appendix, we derive equation~\eq{4order} for velocity
covariances.
Note that particle velocities satisfy equation of motion~\eq{EM}:
\be{EMa}
 \ddot{v}\(\xx\) = \Ds v\(\xx\).
\ee
We introduce covariance of accelerations
\be{}
\zeta = \av{\dot{v}(\xx)\dot{v}(\yy)}.
\ee
Differentiating covariances of velocities~$\kappa$ and covariances of
accelerations~$\zeta$ with respect to time and taking into account
equations of motion~\eq{EM}, \eq{EMa}, yields:
\be{z k}
 \ddot{\kappa} = \(\Dxs + \Dys\) \kappa + 2 \zeta,
 \quad
 \ddot{\zeta}  = \(\Dxs + \Dys\) \zeta + 2 \Dxs\Dys\kappa.
\ee
Excluding~$\zeta$ from system~\eq{z k} yields equation~\eq{4order} for velocity
covariances.

\section{Appendix. Approximation of difference operators}\label{long wave}
In the present appendix, we describe series expansion of difference
operators~$\Dxs$, $\Dys$.
We represent the covariance of particle velocities  in the form~$\kappa\(\rr,
\xx-\yy\)$~(see formula~\eq{kappa z}).
Consider the following expression
\be{Dsxxx}
\begin{array}{l}
\DS \Dxs \kappa\(\rr, \xx-\yy\)
= \omega_*^2
\suma \ba \kappa\(\rr+\frac{1}{2}\va, \xx-\yy + \va\).
\end{array}
\ee
Assume that function~$\kappa$ slowly changes  with the first argument at
distances of order of~$|\va|$. Then series expansion in the right side of
formula~\eq{Dsxxx} with respect to~$\va$ yields
\be{dxsapprox}
\begin{array}{l}
\DS \Dxs \kappa
\approx \omega_*^2\suma \ba \Sa \kappa + \frac{\omega_*^2}{2} \suma \ba \Sa
\va\cdot\nabla\kappa
=
\(\Ds +  \Rs \cdot \nabla \) \kappa,
\\[4mm]
\DS \Rs  = \frac{\omega_*^2}{2} \suma \ba \Sa \va,
\qquad
\Ds = \omega_*^2 \suma \ba \Sa,
\qquad
\Sa \kappa = \kappa\(\rr, \xx-\yy + \va\),
\end{array}
\ee
where~$\nabla = \frac{\partial}{\partial \rr}$ is nabla-operator.
Formulas~\eq{dxsapprox} yield~$\Dxs  \approx \Ds + \Rs \cdot \nabla$.
Analogously we show that~$\Dys  \approx \Ds - \Rs \cdot \nabla$.
Then
\be{Lx-Ly}
\Dxs - \Dys  \approx  2\Rs \cdot \nabla, \qquad  \Dxs + \Dys  \approx  2 \Ds.
\ee
Substitution of the expressions~\eq{Lx-Ly} into equation~\eq{4order}, yields
formula~\eq{4order cont}.

\section{Appendix. Group velocity}\label{c_f general}
In this appendix, we  prove that~$\Vect{c}$, defined by formula~\eq{group vel general},
coincides with the group velocity for the lattice.

The discrete Fourier transform in $d$-dimensional space for an infinite lattice is defined as follows
\be{}
 \begin{array}{l}
 \DS  \hat{\kappa}\(\kk\)= \Phi\(\kappa\) =
 \sum_{j=1}^d\sum_{z_j=-\infty}^{+\infty} \kappa(\zz) e^{-i \kk \cdot \zz},
 \qquad 
\DS \kappa(\zz) = \frac{1}{(2\pi)^d} \int_{-\pi}^{\pi}
\hat{\kappa}(\kk) e^{i \kk \cdot \zz} {\rm d} p_1 ... {\rm d} p_d,
\\[4mm]
 \DS \kk = \frac{1}{a}\sum_{j=1}^d p_j\tilde{\ve}_j,
 \quad
 \zz = \xx-\yy = a \sum_{j=1}^d z_j\ve_j,
  \quad
  \ve_j\cdot\tilde{\ve}_k = \delta_{jk}.
\end{array}
\ee
Here~$a$ is equilibrium distance,  $i$ is the imaginary unit, $\ve_j$ are basis
vectors for the lattice, $\tilde{\ve}_k$ are  vectors of the reciprocal basis,
$\delta_{jk}$ is the Kronecker delta. The discrete
Fourier transform has the following property:
\be{shift}
 \Phi\(\kappa(\zz + \va)\) = \Phi\(\kappa(\zz)\) e^{i \kk \cdot \va}.
\ee
Using the identity~\eq{shift}, we show that
\be{hat Ds gen}
\begin{array}{l}
 \DS
 \Phi\(\Ds \kappa\) = \hat{\Ds} \hat{\kappa},
 \qquad
 \hat{\Ds} =  \omega_*^2 \sum_{\alpha} \ba e^{i \kk \cdot \va},
 \qquad
 \DS
 \Phi\(\Rs \kappa\)  = \hat{\Rs}\hat{\kappa},
  \qquad
  \hat{\Rs}
  =  \frac{\omega_*^2}{2} \sum_{\alpha} \ba \va e^{i \kk \cdot \va}.
\end{array}
\ee

Consider the discrete Fourier transform of equation~\eq{4order cont}.
Calculating the transform using identities~\eq{hat Ds gen}, we obtain
\be{}
\ddddot{\kaph} - 4\hat{\Ds}\ddot{\kaph}  + 4\(\hat{\Rs}\cdot \nabla\)^2\kaph = 0.
\ee
Then vector $\Vect{c}$ is introduced as
\be{c RsDs}
 \Vect{c} = \frac{{\rm Im}\hat{\Rs}}{\sqrt{-\hat{\Ds}}}.
\ee
It can be shown that~$\hat{\Ds}$ and  $\hat{\Rs}$ are related by the following
formula:
\be{Rs Ds}
 \hat{\Rs} = -\frac{i}{2}\frac{{\rm d} \hat{\Ds}}{{\rm d} \kk}.
\ee
Substituting formula~\eq{Rs Ds} into formula~\eq{c RsDs}, we
obtain
\be{c R}
 \Vect{c} =
 \frac{{\rm d} \sqrt{-\hat{\Ds}}}{{\rm d} \kk}.
\ee

Consider the dispersion relation for the lattice.
Substituting~$u(\xx) = A \exp(i (\omega t + \kk \cdot \xx))$ into the equation
of motion~\eq{EM}, yields:
\be{disprel}
  \omega^2 = -\omega_*^2\sum_{\alpha} \ba e^{i \kk \cdot \va} = -\hat{\Ds}.
\ee
Combining the dispersion relation~\eq{disprel} with formulas~\eq{hat Ds gen},
\eq{c R}, yields
\be{cgroup}
  \Vect{c} =  \frac{{\rm d} \omega}{{\rm d} \kk}.
\ee
Therefore, $\Vect{c}$ is equal to the group velocity.

\end{document}